\documentclass[showkeys,showpacs,amssymb,eqsecnum,preprintnumbers,nofootinbib,
superscriptaddress]{revtex4}

\usepackage{graphicx}
\usepackage{latexsym}
\usepackage{amsfonts}
\usepackage{url,hyperref}
\usepackage{bm}

\begin{document}

\newcommand{\vp}{\varphi}
\newcommand{\nn}{\nonumber\\}
\newcommand{\beq}{\begin{equation}}
\newcommand{\eeq}{\end{equation}}
\newcommand{\bed}{\begin{displaymath}}
\newcommand{\eed}{\end{displaymath}}
\def\bea{\begin{eqnarray}}
\def\eea{\end{eqnarray}}
\newcommand{\veps}{\varepsilon}


\title{Graviton emission from a higher-dimensional black hole}
\author{A.~S.~Cornell}
\email[Email: ]{alanc@yukawa.kyoto-u.ac.jp} 
\affiliation{Yukawa Institute for Theoretical Physics, Kyoto University, 
Kyoto 606-8502, Japan}
\author{Wade~Naylor}
\email[Email: ]{naylor@se.ritsumei.ac.jp}
\affiliation{Department of Physics, Ritsumeikan University, 
Kusatsu, Shiga 525-8577, Japan}
\author{Misao~Sasaki}
\email[Email: ]{misao@yukawa.kyoto-u.ac.jp}
\affiliation{Yukawa Institute for Theoretical Physics, Kyoto University, 
Kyoto 606-8502, Japan}

\begin{abstract}
We discuss the graviton absorption probability (greybody factor) and
the cross-section of a higher-dimensional Schwarzschild black hole
(BH). We are motivated by the suggestion that a great many BHs may be produced at the LHC and bearing this fact in mind, for simplicity, we shall investigate the intermediate energy regime for a static Schwarzschild BH. That is, for
$(2M)^{1/(n-1)}\omega\sim 1$, where $M$ is the mass of the black hole and $\omega$ is the energy of the emitted gravitons in $(2+n)$-dimensions. To find easily tractable solutions we work in the limit $l \gg 1$, where $l$ is the angular momentum quantum number of the graviton.
\end{abstract}

\begin{figure}[t]
\vspace{-1.4cm}
\hspace{-16.25cm}
\scalebox{0.085}{\includegraphics{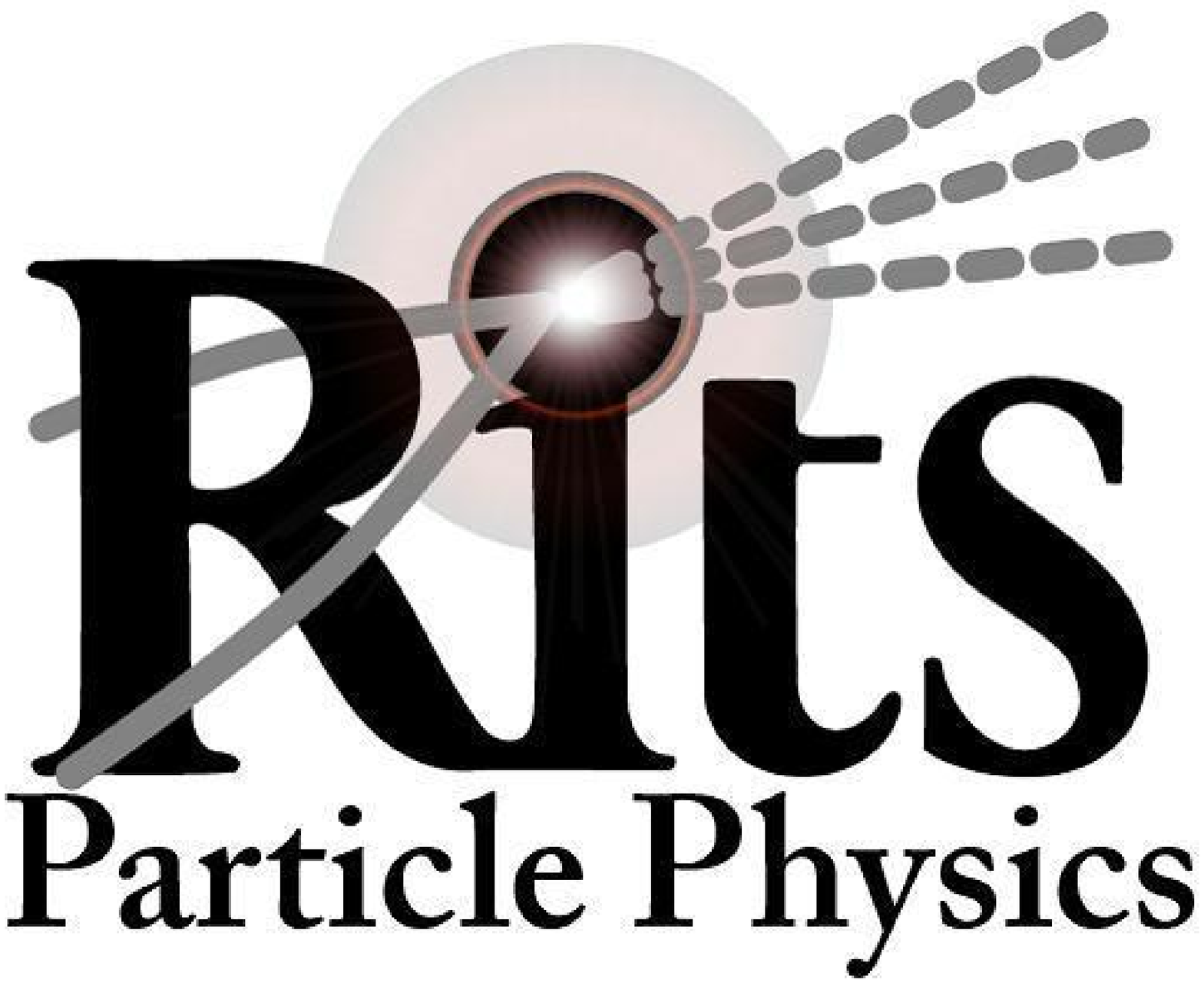}} 
\end{figure}

\pacs{0470.Dy, 11.10.Kk}
\keywords{Extra dimensions, Black Holes}
\preprint{YITP-05-59}
\preprint{RITS-PP-006}
\date{October 2, 2005}
\maketitle

\section{Introduction}

\par Much discussion has recently focused on the emission rates of TeV BHs as motivated by the proposition that the quantum gravity scale can be brought down to as low as a TeV in some higher-dimensional BH models \cite{GT,DL,HHBS}. Most cases have focused on the low-energy emission of scalar or spinor fields from higher-dimensional Schwarzschild and slowly rotating
Kerr BHs, for a review see reference \cite{Krev}. Numerical methods also allow us to evaluate the emission in the full energy range \cite{Krev}. 

\par The importance of bulk emission of gravitons, as well as possible recoil effects, was highlighted recently in references \cite{FS1,FS2} (also see references \cite{Stoj,FT}).
However, little attention has been paid to the actual emission of gravitons from a BH (though there has been some work relating to the quasi-normal modes (QNMs) of higher-dimensional BHs \cite{Card1,Card2,Konop,Konop2,Berti}) and furthermore it is useful to have analytic
expressions for the cross-sections etc. not just in the low-energy or
high-energy classical regime. In this article we shall discuss BH
cross-sections for gravitons in what we shall call the intermediate energy
regime for the variable $\varepsilon=2M\omega^{n-1}$, that is where
$\varepsilon\sim 1$. In this regime the energy of the particle is near the
peak of the potential barrier in the associated scattering problem. The
classical cross-section is reproduced in the high-energy limit,
$\varepsilon\gg 1$, as we shall discuss later. 

\par In this article we shall focus on the static Schwarzschild BH, though in some models a rotating BH does not necessarily spin-down to zero, but evolves to a non-zero angular momentum \cite{Chambers,NYTM}. However, as yet the gravitational perturbation equations for a higher-dimensional rotating BH are unknown. Thus, we shall investigate the intermediate energy regime for the static case, hoping that it may encode some of the properties of the rotating case. Indeed, graviton super-radiance for a large rotation parameter is expected in the intermediate energy regime. To this end only
recently has the case of spin-zero emission for the high-energy
and high angular momentum regime been studied \cite{Harris:2005jx,Duffy:2005ns}, where numerical methods were employed. Note that scalar emission for low-energies and low angular momentum has been studied in references \cite{IOP2,Frolov}.

\par The particular background we shall investigate is that of a static
Schwarzschild BH in $(2+n)$-dimensions. To perform the calculation we
shall recall some results recently derived for the gravitational
perturbations of a higher-dimensional maximally symmetric BH
\cite{KI}. As mentioned earlier, although it seems likely that TeV 
BHs will probably be highly rotating when produced \cite{NYTM}
there is not yet a method for separating the gravitational perturbations
on a higher-dimensional Kerr background. Thus, we shall content ourselves
with the static Schwarzschild case for now.

\par Gravitational perturbations in general separate into scalar
(polar) and vector (axial) perturbations. In more than four-dimensions
there is an extra degree of freedom corresponding to tensor
perturbations. As shown in reference \cite{KI} we have for the scalar
perturbations;\footnote{Note that for $n=2$ the scalar
(polar) perturbation agrees with the Zerilli  equation \cite{Zer}.}
\beq
-f\frac{d}{dr}\left(f\frac{d\Phi}{dr}\right)+V_S\Phi
 =\omega^2\Phi~,
\label{sme}
\eeq
where 
\beq
f(r)=1-{2M\over r^{n-1}}
\label{eff}
\eeq
and
\beq
V_S(r)=\frac{f\,H(r)}{16r^2 [m+\frac 1 2n(n+1)(1-f)]^2} ,
\label{scalarpot}
\eeq
with
\bea
H(r)&=&\Big(n^4(n+1)^2(1-f)^3
+n(n+1)\left[4(2n^2-3n+4)m+n(n-2)(n-4)(n+1)\right](1-f)^2\nn
   & &\hspace{1cm} - 12n\left[(n-4)m+n(n+1)(n-2) \right]m(1-f)
   +16m^3+4n(n+2)m^2~
\Big) .
\eea
In the above
\beq
m=l_S(l_S+n-1)-n , \qquad l_S=2,3,\,\dots
\eeq
and as discussed in reference \cite{KI}
the mode $l=0$ corresponds to adding a small mass to the BH; the mode $l=1$ is a pure gauge mode. 
Note that the true mass of the BH, $M_{\rm BH}$, as measured on the brane, is the same as that in the bulk \cite{Tang,MP} and hence;
\beq
   M_{\rm BH}= \frac{ n {\cal A}_{n} M }{8 \pi c^2 G_{n+2}}~,  
\eeq
where 
\beq
{\cal A}_n = {2\pi^{(n+1)/2}\over\Gamma[(n+1)/2]}
\eeq
is the area of a unit $n$-sphere, $G_{n+2}$ is the $(2+n)$-dimensional
Newton constant, and $c$ is the speed of light. In what follows we shall
set $G_{n+2}=c=1$.

\par The vector and tensor perturbations have a much simpler form, which
can be written in one concise equation as;
\beq
-f \frac{d}{dr}
\left( f\frac{d \Phi}{dr} \right)
+V_{V/T}\Phi=\omega^2\Phi~,
\label{vtme}
\eeq
with\footnote{The vector (axial) perturbations for $n=2$ reduces to the
standard Regge-Wheeler equation \cite{RW}.}
\beq
V_{V/T}(r)=\frac{f}{r^2}\left(l(l+n-1)+\frac{n(n-2)}{4}
-{\mu_{V/T}\over4} \frac{n^2M}{r^{n-1}}\right)
\qquad\qquad\qquad
\begin{array}{c}
l_{V}=2,3,\,\dots 
 \\
l_T=1,2,\,\dots 
\end{array}
\label{vectenpot}
\eeq
where $\mu_V=3$ for the vector perturbations, while for the tensor
perturbations $\mu_T=-1$. Note that the mode $l=1$ of the vector perturbation, which is absent from the above spectrum, represents a purely rotational mode of the BH. 

\par We shall mainly be interested in evaluating the scattering
cross-section, which is related to the absorptions probability, $|{\cal
A}_{lP}(\omega)|^2$, by \cite{CCG}; 
\beq
\sigma(\omega) = C{(4\pi)^{(n-1)/2}\over
\omega^{n}}\,\Gamma\left({n+1\over 2}\right)\,
\sum_l^\infty \sum_P D_{lP}|{\cal A}_{lP}(\omega)|^2=\sigma_S+\sigma_V+\sigma_T ~,
\label{cross}
\eeq
where $P$ denotes each respective perturbation and the graviton 
normalization is \cite{CCG};
\beq
C={2\over (n+2)(n-1)}~.
\eeq
The degeneracy of each perturbation, $D_{lP}$, is \cite{Rubin};
\bea
\label{deg}
&&D_{lS}=\frac{(2l+n-1)\,(l+n-2)!}{(n-1)!\, l !}~,
\nn
&&D_{lV}=\frac{l(l+n-1)(2l+n-1)(l+n-3)!}{(n-2)!\, (l+1) !}~,
\nn
&&D_{lT}=\frac{(n+1)(n-2)(l+n)(l-1)(2l+n-1)(l+n-3)!}{2(n-1)!(l+1)!}~,
\eea
for a given angular momentum channel $l$, which is the spin-2 generalization of the result given in reference \cite{Krev}. 
Thus, only scalar and vector perturbations contribute for $n=2$ and the partial sums effectively start from $l=2$, even for the tensor perturbation. 

\section{Intermediate energy approach}

\par To evaluate the absorption probability and hence the graviton
emission rate in the intermediate energy regime we shall use the WKB
approach of Iyer and Will \cite{IW}. This was recently used in the
higher-dimensional context by Berti {\it et al.} \cite{Berti} (also see reference \cite{Card2}) to
investigate the gravitational energy loss of high-energy particle
collisions using a QNM analysis. In the following we shall work to lowest
order in the generalised WKB method of reference \cite{IW}; however, to check the validity of the method we go up to second order to verify that the next order correction is small.

\par As discussed in reference \cite{Berti}, in order to use the WKB method we must
rewrite the perturbations in the $(n+2)$-dimensional tortoise coordinate
defined by;
\beq
\frac{dr_*}{dr}=\frac{1}{f(r)} ,
\label{oveff}
\eeq
which implies that
\beq
r_*=r+\frac{2M}{n-1}\sum_{j=0}^{n-2}\frac{\ln(r/\alpha_j-
1)}{\alpha_j^{n-2}}\,,
\eeq
where
\beq
\alpha_j=(2M)^{1/(n-1)}{\rm e}^{2\pi{\rm i}j/(n-1)} \qquad
(j=0,\dots,n-2)\,.
\eeq
The tortoise coordinate, $r_*$, given above becomes quite complicated in
more than four-dimensions and in our case it will be more convenient to
work with the original coordinate $r$ and use equation (\ref{oveff}) to convert
derivatives. Thus the perturbation equations reduce to the standard 
Schr\"{o}dinger form;
\beq
\left(\frac{d^2}{dr_*^2}+Q_P(r_*)\right)\Phi=0\, ,
\qquad\qquad Q_P(r_*)=\omega^2-V_P(r_*).
\eeq
The subscript $P$ denotes any one of the three possible
perturbations. Note that the potential, equation (\ref{scalarpot}), is defined in
terms of $r$ and not $r_*$, the tortoise coordinate.

\par As discussed in reference \cite{IW}, an adapted form of the WKB method can be
employed to find the QNMs or the absorption probability, which we are
primarily interested in, when the scattering takes place near the top of
the potential barrier. In the following we shall use the same notation as reference \cite{IW}. The absorption probability, up to second order in the WKB
expansion, is found to be;
\beq
|{\cal A}_l(\omega)|^2={1\over 1+e^{2i\pi(\nu+1/2)}}~,
\label{abprob}
\eeq
where 
\beq
\nu+1/2=i(2 Q_{P_0}'')^{-1/2} Q_{P_0}-\Lambda
\label{nu}
\eeq
and 
\beq
\Lambda
=i(2 Q_{P_0}'')^{-1/2}\left(\left[\frac 3 8 
\left({Q^{(4)}_{P_0}\over 12 Q_{P_0}''}\right)- 
\frac{7 }{32} \left({Q^{'''}_{P_0}\over 3 Q_{P_0}''}\right)^2
\right]
+(\nu+1/2)^2\left[\frac 6 4 \left({Q^{(4)}_{P_0}\over 12 Q_{P_0}''}\right) - 
\frac{30 }{16} \left({Q^{'''}_{P_0}\over 3 Q_{P_0}''}\right)^2
\right]\right)\,,
\label{lambda}
\eeq
these being parameters determined in reference \cite{IW}, and where the subscript zero
denotes the maximum of $-\,Q(r_*)$.

\par By substituting equation (\ref{lambda}) into equation (\ref{nu}) we can eliminate
$\Lambda$  to obtain an expression solely in terms of $\nu+1/2$; however,
because we wish to only know the size of the second order correction we
can eliminate $\nu+1/2$ to find $\Lambda$;
\beq
\Lambda=\frac1{2C}\left(1+2AC\pm \sqrt{1+4(A-B)C}~\right)\,,
\label{secord}
\eeq
where
\beq
A=i(2 Q_{P_0}'')^{-1/2}Q_{P_0} , 
\eeq
and
\beq
B=
i(2 Q_{P_0}'')^{-1/2}\left[\frac 3 8 
\left({Q^{(4)}_{P_0}\over 12 Q_{P_0}''}\right)- 
\frac{7 }{32} \left({Q^{'''}_{P_0}\over 3 Q_{P_0}''}\right)^2
\right] , 
\qquad\qquad
C=i(2 Q_{P_0}'')^{-1/2}\left[\frac 6 4 \left({Q^{(4)}_{P_0}\over 12 Q_{P_0}''}\right) - 
\frac{30 }{16} \left({Q^{'''}_{P_0}\over 3 Q_{P_0}''}\right)^2
\right]~.
\label{BC}
\eeq
In general it is just a simple but tedious exercise in algebra to evaluate
the second order correction. However, as we discuss in the next section, 
by considering the potential to leading order, ${\cal O}(l^0)$, we can estimate the
size of the second order correction without making any explicit
calculations.

\section{Zeroth-order approximation}

\par To illustrate the method as simply as possible let us work to zeroth
order in the WKB expansion for the case of large $l$. However, in order to find a difference between the respective perturbations we should
consider up to $O(l^0)$ in the potential, which in this case is given by;
\beq
V_S(r)\approx{f\over r^2} \left(l^2+(n-1)l +\frac{n(n-2)}{4}+{n\over4}(8-7n)(1-f)\right)\,, 
\label{scallrg}
\eeq
for the scalar perturbation and 
\beq
V_{V/T}(r)=\frac{f}{r^2}\left(l^2+(n-1)l+\frac{n(n-2)}{4}
-\frac{\mu }{4}n^2(1-f)\right)\, ,
\label{vectenlrg}
\eeq
for the vector and tensor potential, which is an exact expression to ${\cal O}(l^0)$, see equation (\ref{vectenpot}). Importantly, we
can immediately notice that in four-dimensions, $n=2$, the scalar and
vector potentials are identical to ${\cal O}(l^0)$. The equivalence of the scalar
and vector potentials for general $l$ is well known in four-dimensions
\cite{Chandra}, however, as discussed in reference \cite{KI}, this is not true for
the case of general dimensions.\footnote{To ${\cal O}(l)$ all the perturbations
become identical, regardless of the dimension.} This will have important
consequences for the graviton emission of a higher-dimensional BH,
which we shall discuss later. For convenience let us write a general
formula for all the perturbations, valid up to $O(l^0)$;
\beq
V_P(r)=\frac{f(r)}{r^2}\left(\hat l^2+\beta
-\frac{\alpha_P }{4}(1-f)\right) , \qquad \mathrm{where} \qquad 
\hat l^2= l^2 + (n-1)l\qquad\qquad \forall~l=2,3,\,\dots~.
\label{genpot}
\eeq
In the previous equation we have used;
\beq
\beta={n(n-2)\over 4} , 
\eeq
and
\beq
\label{coeff}
\alpha_S=n(7n-8) , \qquad\qquad\alpha_V=3n^2 , 
\qquad\qquad\alpha_T=-n^2~.
\eeq

At this stage it will be convenient to change variables to $z=\omega r$,
for example see reference \cite{PS}, and by doing so the WKB equation now
becomes;
\beq
\left(\frac{d^2}{dz_*^2}+Q_P(z_*)\right)\Phi=0\, , 
\qquad\qquad Q_P(z_*)=1-V_P(z_*)\,,
\eeq
where
\beq
V_P(z)=\frac{f(z)}{z^2}\left(\hat l^2+\beta
-\frac{\alpha_P }{4}{\varepsilon\over z^{n-1}}\right) , 
\eeq
and 
\beq
f(z)=1-{\varepsilon\over z^{n-1}}\,;\qquad\qquad 
\veps=2M \omega^{n-1}\,.
\eeq
The expression for the absorption probability, equation (\ref{abprob}), is defined
explicitly in terms of derivatives of the potential with respect to the
tortoise coordinate $z_*$ (or $r_*$, where $z_*$ has been
defined in an analogous way to $r_*$); however, in terms of $z$, up to
second order in derivatives of the potential, we find;
\bea
{dQ\over dz_*}&=&{dz\over dz_*} {dQ\over dz}=f(z){dQ\over dz}
=\left(1-{\varepsilon\over z^{n-1}}\right){dQ\over dz}~,
\\
{d^2Q\over dz^2_*}&=& f(z){d\over dz}\left(f(z){dQ\over dz}\right)
=\left( 1 - {\epsilon\over z^{n-1} }  \right) \,
  \left(z^{-n}\left( -1 + n \right) \,\epsilon \,{dQ\over dz} 
+ \left( 1 - {\epsilon\over z^{n-1}}  \right) \,{d^2Q\over dz^2} \right) ~. 
\label{2ndder}
\eea
Note that at the maximum of the potential, $z_0$, the first term in equation
(\ref{2ndder}) is zero and thus simplifies our calculation. 

Hence, from equation (\ref{nu}) with $\Lambda=0$, we are easily lead to
the result;
\beq
|{\cal A}_{lP}|^2={1\over 1+e^{ {2\pi\over K}
\big[ f(z_0)\left(\hat l^2 + \beta  - \frac\alpha{4}\,  \varepsilon{{z_0}}^{1 - n} \right)-z_0^2        
\big] 
}}\,,
\label{ablrgl}
\eeq
where we have written the general perturbation as;
\bea
K(\hat l,z_0,\varepsilon) &=&{
(z_0^n-\varepsilon\,z_0  )
\over{\sqrt{2}}\,{{z_0}}^{2\,n }
}{\sqrt{-2\,n\,\left( 1 + 2\,n \right) \,\alpha \,{\epsilon }^2\,
      {{z_0}}^2 -24\,\left(\hat l^2 + \beta  \right) \,{{z_0}}^{2\,n} +
     \left( 2 + 3\,n + n^2 \right) \,
      \left( 4\,\hat l^2 + \alpha  + 4\,\beta  \right) \,\epsilon \,
      {{z_0}}^{1 + n}}}\,
      \nn
\eea
and we have dropped the subscript $P$ for clarity. Choosing the appropriate $\alpha_P$ from equation (\ref{coeff}) then determines the absorption probability for each respective
perturbation mode. The location of the maximum, $z_0$, for the potential
$-\,Q_P(z_0)$ can also be solved, see appendix A. This also depends on the
type of perturbation. 

\par Before we proceed any further we should verify that the second order
correction is small as compared to the zeroth order one. Given that $V_P$
is of order ${\cal O}(l^2)$, see equation (\ref{genpot}), and that $l$ is
independent of $r$ or $r_*$, it is simple to see that;
\beq
i(2 Q_{P_0}'')^{-1/2}\sim {\cal O}(l^{-1})~.
\eeq
The terms in the square brackets for $B$ and $C$, see equation (\ref{BC}),
are of ${\cal O}(l^0)$ and hence (using the relation above) $B$ and $C$
contribute to order ${\cal O}(l^{-1})$. Furthermore the term $A$ is of order
${\cal O}(l)$. Thus the second order contribution, $\Lambda$, see
equation (\ref{secord}), is of order ${\cal O}(l^{-1})$ and is therefore
negligible in the limit $l \gg 1$. 
This is also true for the high-energy limit $\varepsilon \gg 1$, as well as for intermediate energies $\varepsilon \sim 1$.\footnote{The WKB method breaks down for the low-energy case $\varepsilon \ll 1$.}

\section{Results}

\par In this section we shall present our results as based on the analytic expressions derived in the previous section. Furthermore we shall also demonstrate that the geometric optics limit is reproduced for large values of $\varepsilon$. Note that for completeness we shall also derive this expression.

\par
\begin{figure}[t]
\begin{center}
\scalebox{0.85}{\includegraphics{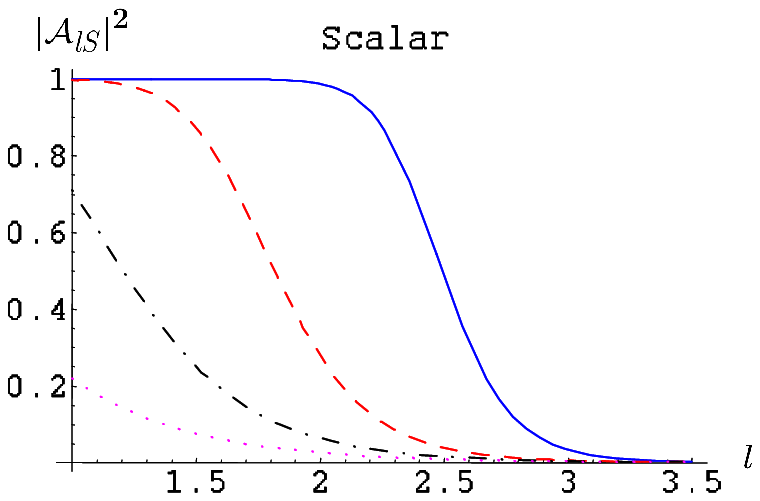}} 
\hspace{1.5cm}
\vspace{1.35cm}
\scalebox{0.85}{\includegraphics{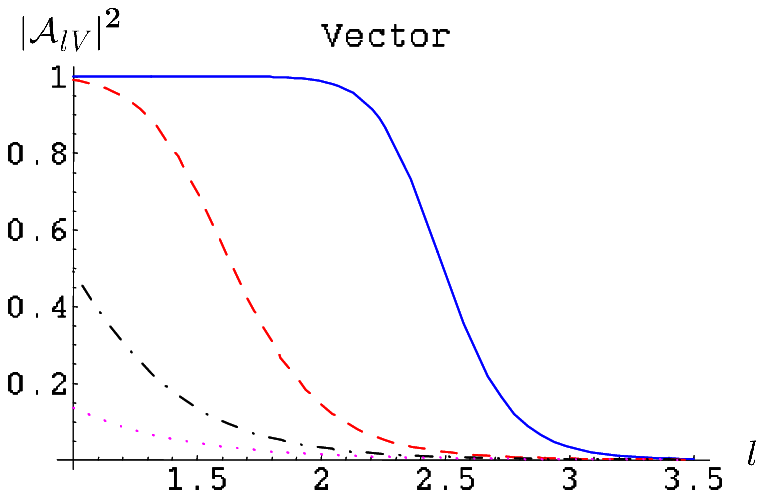}}
\scalebox{0.85}{\includegraphics{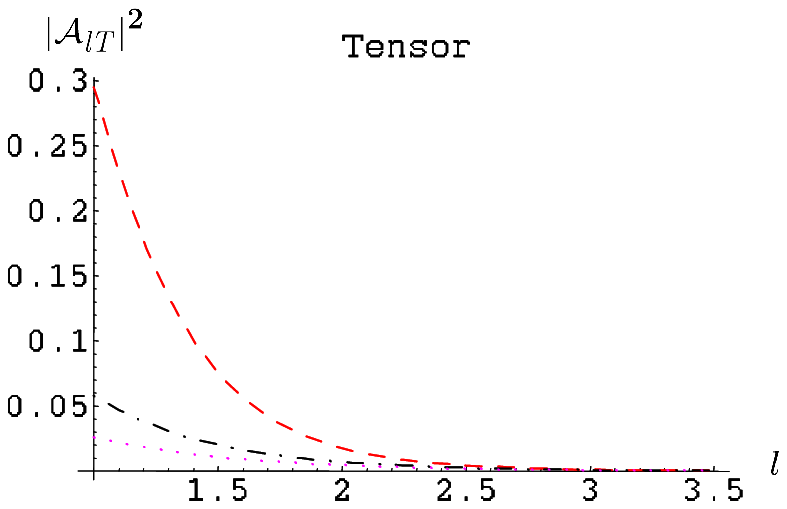}}
\end{center} 
\caption{Black hole absorption probability, $|{\cal A}_{lP}|^2$, as a
function of angular momentum $l$ (extended to real numbers) for $\varepsilon=1$.
Blue (solid), red (dashed), black (dot-dashed) and purple (dotted) curves correspond to $n=2,3,4$ and $5$ respectively. Note that the scalar and vector perturbations are only valid from $l=2$ onwards, however, for comparison with the tensor case we start 
at $l=1$. 
\label{absplot}}
\end{figure}

 We have presented our first set of results in Fig.~\ref{absplot} where we have plotted the absorption probability as a function of $l$ for each of the perturbations. We have also considered scenarios involving different dimensions, $n$, for $\varepsilon\sim 1$. Firstly, the scalar mode has the largest
contribution, followed by the vector and then the tensor
perturbations; however, this does depend on the value of
$\varepsilon$. Secondly, we see that the absorption probability is larger
for smaller $n$. In fact numerical plots for different $\varepsilon$ show
that $|{\cal A}_{lP}|^2$ saturates to unity for larger and larger $l$ as
$\varepsilon$ increases. This implies that larger $l$ is required in the
angular momentum sum for the cross-section, when obtaining the geometric
optics limit, see below.

\begin{figure}[t]
\begin{center}
\scalebox{1.1}{\includegraphics{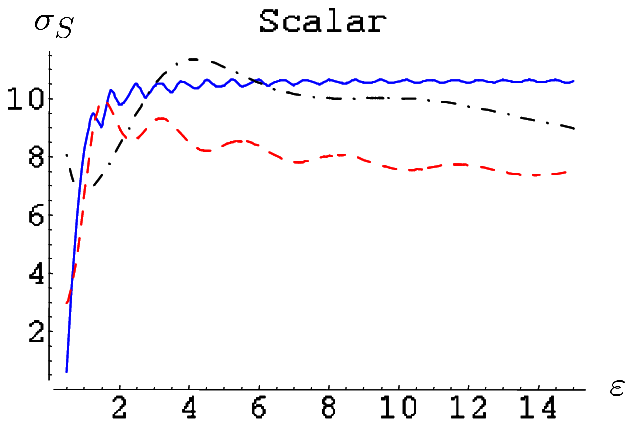}} 
\vspace{0.75cm}
\hspace{1.75cm}
\scalebox{1.1}{\includegraphics{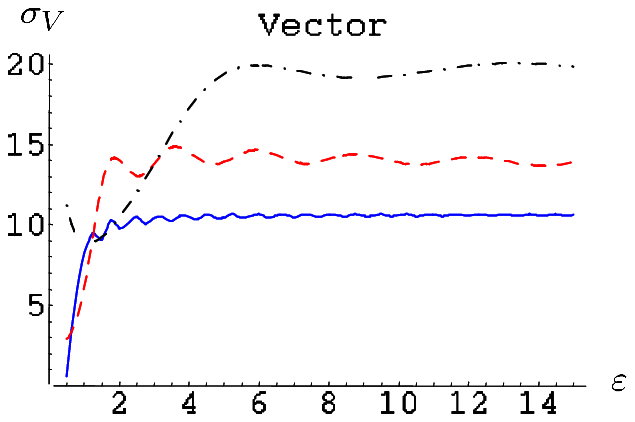}}
\scalebox{1.1}{\includegraphics{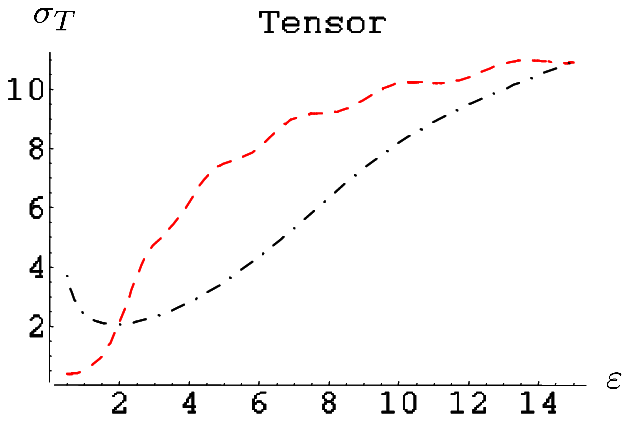}}
\hspace{1.75cm}
\scalebox{1.1}{\includegraphics{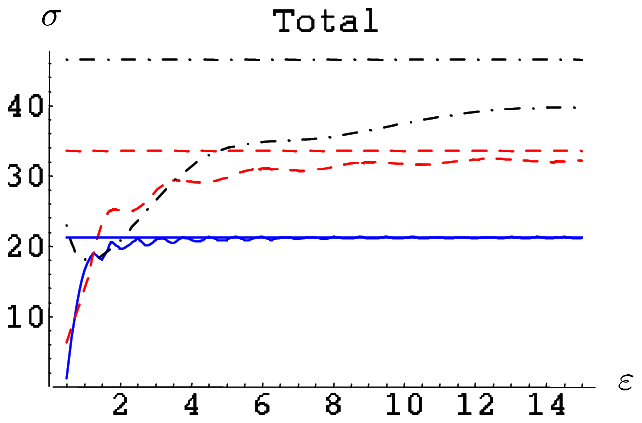}}
\end{center} 
\caption{Black hole cross-sections as a function of $\varepsilon$, in units
of $(2M)^{n\over n-1}$, for each gravitational perturbation and the total one. The
horizontal lines correspond to the classical cross-section. Blue (solid), red (dashed) and black (dot-dashed) curves correspond to $n=2,3$ and $4$ respectively. Note that there is no tensor perturbation for the $n=2$ case.
\label{xplot}}
\end{figure}

Turning to the cross-section, as given by\footnote{The cross-section effectively starts from $l=2$ for the tensor perturbations, like for the other perturbations, because of the form of the degeneracy $D_{lT}$.};
\beq
\sigma(\omega) = C\,(4\pi)^{(n-1)/2}\,\Gamma\left({n+1\over 2}\right)\,
\left({2M\over\varepsilon}\right)^{n\over n-1}
\sum_{l=2}^\infty \sum_P{D_{lP}\over 1+e^{ {2\pi\over K}
\big[ f(z_0)\left(\hat l^2 + \beta  - \frac\alpha{4}\,
\varepsilon{{z_0}}^{1 - n} \right)-z_0^2        
\big] 
}}~,
\label{xsec}
\eeq
where $D_{lP}$ is given in equation (\ref{deg}) and remembering that 
$\alpha=\alpha_P$. It should be noted that though we are primarily
interested in the intermediate energy regime, $\varepsilon \sim 1$, with
large angular momentum, $l\gg 1$, the above expression can also be applied
to the case where $l>\varepsilon^{1/(n-1)}\sim 1$. Note also that although the next order
correction in the WKB expansion becomes larger for $l\sim {\cal O}(1)$ we can still
extrapolate the results hoping the errors are not too large. In
Fig.~\ref{xplot} we plot the cross-section, equation (\ref{xsec}), as a function of $\varepsilon$ and to ensure convergence in the partial wave sums we sum up to 
$l_{max}\sim 3\,\varepsilon_{max}$. The total cross-section tends to the classical one for large $\veps$ as we shall now explain.

\par In the intermediate energy approach (for large $l$) the 
absorption probabilities satisfy $A_{lS}=A_{lV}=A_{lT}$, see Fig.~\ref{absplot}, and  the degeneracies satisfy \cite{CCG};
\beq
D_{lS}\approx C\left( D_{lS} + D_{lV}+D_{lT}\right)~.
\label{appdeg}
\eeq
Furthermore, the classical cross-section corresponds to the high-energy limit,
$\varepsilon^{1/(n-1)}\gg l \gg 1$, which then implies the absorption coefficients $A_{lP}\sim{\cal O}(1)$ and in this limit the sum over $l$ has a cut-off at $l\approx b\,\omega$ (for example, see reference \cite{DeWitt}) where $b$ is the critical radius (or
impact parameter) at which the BH ceases to absorb radiation. This is given by equation (12) of reference \cite{EHM}, for a massless/relativistic particle as;
\beq
b=\left(n+1\over 2\right)^{1\over n-1}\left(n+1\over n-1\right)^{1\over
2}(2M)^{1\over n-1}~.
\eeq
Given the cut-off in the mode sum we have;
\beq
\sum_l^{b\omega} D_{lS}\sim {2(b\omega)^n\over n!}
\eeq
for large $\omega$, and substituting this into equation (\ref{xsec}), using (\ref{appdeg}), leads to the classical cross-section;
\beq
\label{go}
\sigma_{c}={(4\pi)^{(n-1)/2}\over
\omega^{n}}\,\Gamma\left({n+1\over 2}\right)
{2(b\omega)^n\over n!} ={2\sqrt\pi(\pi)^{n-1\over2}\over
n \Gamma(n/2)}\left(n+1\over 2\right)^{n\over n-1}\left(n+1\over
n-1\right)^{n\over 2}(2M)^{n\over n-1}~,
\eeq
where in the second step we used the gamma duplication formula. This agrees with the standard result in $(2+n)$-dimensions, compare with reference \cite{Krev} (after making the substitution $r_H^{n-1}=2M$). 

As is well known, the high-energy cross-section is
independent of the particle species, and likewise we see that it is
independent of the graviton cross-section. These are
represented by horizontal lines in the bottom right plot in Fig.~\ref{xplot} and we see that our analytic results correctly reproduce the high-energy limit
$\varepsilon\gg 1$ (note that for higher-dimensions larger values of $\varepsilon$ are required to obtain the geometric optics limit\footnote{The geometric optics limit in equation (\ref{go}) has a maximum at $n\approx 7$, which implies that for $n>7$ (i.e. for greater than 9-dimensions) the classical cross-section starts to decrease.}). 

\section{Conclusion}

\par In conclusion we have shown in this paper that the WKB method of Iyer and Will \cite{IW} can be applied to the case of graviton emission from a Schwarzschild BH. Indeed
our results reproduce the classical cross-section in the high-energy
limit. We have also presented new results for the intermediate energy
regime.

\par In the low-energy limit, $M\omega \ll 1$, our method breaks down as
can be seen from Fig.~2, where the cross-section starts to diverge. It is
well known that the cross-section is proportional to $\omega^{2l+n}$
in $(2+n)$-dimensions \cite{Krev}, which essentially corresponds to $s$-wave scattering for a spin-zero field, as the lowest $l$ modes dominate the cross-section. Indeed it is straightforward to obtain the low-energy cross-section for the vector and tensor perturbations by employing the standard technique of matching the near horizon and far field solutions, for example using the techniques discussed in references \cite{KR,Unruh,alstar}. However, there is a subtlety with the higher-dimensional Zerilli (scalar) equation: In four-dimensions the Regge-Wheeler
(vector) solution is usually used to find the absorption probability,
which is equivalent to that for the Zerilli equation, as the two
solutions are identical in four-dimensions \cite{Chandra}, but as pointed
out in reference \cite{KI}, in higher dimensions no such relation exists. Thus, if
we attempt to use the standard technique of matching the near horizon and
far field solutions directly for the Zerilli equation it appears
that numerical methods seem more amenable. Regardless, the focus of this current work has been the intermediate energy regime given that this is
where we expect graviton emission to be most interesting for a rapidly rotating BH.

\par Although in this article we focused on the static Schwarzschild BH, we can also apply our method to the case of spin-zero field emission from a Kerr BH \cite{CJNS2}. Note that though the solution to the graviton perturbations for a higher-dimensional Kerr BH have not yet been found, we are encouraged by the fact that there are some 
similarities between the total graviton  and spin-0 cross-sections, see Appendix B. As can be seen from Fig.~\ref{bosex}, the total cross-section, which is the sum of the perturbations (scalar, vector and tensor) is of the same order of magnitude as the spin-0 case, i.e. $O(\sigma_S+\sigma_V+\sigma_T)\sim O(\sigma_B)$, and such an approximation may also be valid for the rotating case.

\par The rotating case is particularly interesting due to the phenomenon of super-radiance (for example see reference \cite{alstar}), where the absorption probability becomes negative. Super-radiance has been discussed in the higher-dimensional context recently in references \cite{Harris:2005jx,NYTM,IOP2,Duffy:2005ns,Frolov}. However, the highly rotating case has only been considered using numerical techniques; whereas we expect that our approach allows for an analytic expression for such a case. Furthermore, our approach should also allow one to evaluate the QNMs analytically for the rapidly rotating case, as has recently been done in the limit $a\to\infty$ \cite{CSY}. In the four-dimensional case the rotation parameter $a=1$ is bounded to $a_{\rm max}=1$, for example see reference \cite{Chandra}. This fact has meant that there have been no prior investigations of the QNMs of a Kerr BH with a large rotation parameter. However, in $(2+n)$-dimensions there are $(n+1)/2$ rotation parameters, where for $n>3$ the rotation parameters are unbounded \cite{MP}.\footnote{If we assume that the BH is created on a thin brane then there is only one rotation parameter, for example see reference \cite{Harris:2005jx}.} Indeed, as we mentioned, even after the spin-down phase in some models the BH remains rotating \cite{Chambers,NYTM}.

\par Another interesting case for which the WKB method can be applied is that for emission from a charged (Reisner-Nordstrom) BH. Recently the graviton perturbations have been found in reference \cite{KI2} (along with their QNMs \cite{Konop2}). However, in this case the solutions are complicated by the fact that there are now solutions for the electromagnetic field itself. We hope to present results for this model in the near future.

\section*{Acknowledgments}

\par The work of ASC was supported by the Japan Society for the Promotion of Science (JSPS), under fellowship no. P04764. The work of MS was supported in part by Monbukagakusho Grant-in-Aid for Scientific Research(S) No. 14102004 and (B) No.~17340075.
\\
\par{\it Note added:} After the completion of this article, similar work appeared in \cite{CCG}, which correctly pointed out that the degeneracy factor for the graviton cross-section should be that for a spin-2 field \cite{Rubin}, which is different to the spin-0 case, for $n>2$. Our main results do not change; however, only the total cross-section now coincides with the classical one, due to the normalisation. This actually improves our attempt to model the spin-2 perturbations by a spin-0 field, as is discussed in Appendix B.

\appendix

\section{Location of the maximum, $Q_{P_0}$}
    
\par In order to find the maximum of the potential it is convenient to
work with the coordinate;
\beq
x=1-f={2M\over r^{n-1}}\,,
\label{ex}
\eeq
rather than $r_*$ or $r$. That is we wish to find the roots of;
\beq
0={dV_P\over dr_*}={dr\over dr_*} {dV_P\over dr}=(1-n)(2M)^{1\over 1-n}(1-x)
x^{n\over n-1}~{dV_P\over d x} . 
\eeq
Here we should mention that in our case, even though all the expressions
depend on $Q_P$,  the potential can be written in the form
$Q_P=\omega^2-V_P$ and thus, for a given energy, $\omega$, the maximum
of $-Q_P$ corresponds to the maximum of $V_P$.

\par Therefore, apart from the solutions at the horizon ($x=1$) and at
infinity ($x=0$), we merely require the roots of;
\beq
{dV_P\over d x}=0~.
\label{roots}
\eeq
Working to $O(l)$ in the potential allows the maximum to be easily found;
\beq
x_0^P={2\over n+1}\,,
\label{root}
\eeq
which is a result valid for all three perturbations. However, as we have previously discussed, in order to see a difference between the perturbations we must
work to order $O(l^0)$. Hence, for the gravitational perturbations we
obtain;
\beq
x_0^P=\frac{\left( 1 + n \right) \,\left( 4\,\hat l^2 + \alpha  +
4\,\beta  \right)  - 
    {\sqrt{-64\,n\,\alpha \,\left( \hat l^2 + \beta  \right)  + 
        {\left( 1 + n \right) }^2\,
         {\left( 4\,\hat l^2 + \alpha  + 4\,\beta
\right) }^2}}}{4\,n\,\alpha }\,,
\eeq
where, as before, we are using the shorthand notation $\hat
l^2=l^2+(n-1)l$ and the coefficients $\alpha$ and $\beta$ (ignoring the
subscripts) as defined in equation (\ref{coeff}). Note that if we 
expand the square root in powers of large $l$ we obtain equation
(\ref{root}).

\par The location of the maximum in terms of the coordinates $r$ and $z$
is then simply given by;
\beq
r_0^P=\left(2M\over x_0^P\right)^{1\over n-1}
\qquad \mathrm{and} \qquad
z_0^P=\omega r_0^P=\left(\varepsilon\over x_0^P\right)^{1\over n-1}\,,
\label{maxr}
\eeq
which is independent of the perturbation up to $O(l)$, see
equation (\ref{root}).

\section{Spin-Zero Emission}

\par In this appendix we briefly discuss the case of emission of a bulk
scalar (spin-zero) field. As we shall see the effective potential is
equivalent to that for the tensor perturbations. Following
reference \cite{KR}, after a separation of variables, the scalar radial
equation, $g^{\mu\nu}\nabla_\mu \nabla_\nu\,\phi(x)=0$, is found to be;
\beq
\frac{f(r)}{r^{n}}\,\frac{d \,}{dr}\,
\biggl[\,f(r)\,r^{n}\,\frac{d R(r)}{dr}\,\biggr] +
\biggl[\,\omega^2 - \frac{f(r)}{r^2}\,l\,(l+n-1)\,\biggr] R(r) =0 \,.
\label{rad}
\eeq
This equation can be written in the WKB form by applying the tortoise
coordinate defined in equation (\ref{oveff})\footnote{In reference \cite{KR} a slightly different choice of tortoise coordinate is made.} and rescaling
the radial solution by $R(r_*)=r^{-n/2}\Phi(r_*)$. A short calculation leads to;
\beq
\frac{d^2 \,\Phi^2}{dr_*^2}+\omega^2\,\Phi(r_*)
 -\frac{f(r)}{r^2}\left[
l(l+n-1)+{n\over4}(n-2)+{n^2\over 4}(1-f)\right]\Phi(r_*) = 0 , 
\qquad\qquad l_B=0,1,2,\dots
\label{WKB}
\eeq

One can then easily verify that this equation is identical to that for the
tensor perturbations of the graviton, see equation (\ref{vectenpot}); however, the degeneracy factor, $D_{lP}$, will be different. The
only difference between the two being that the sum over $l$ starts from $l=0$
for a spin-zero field, not $l=2$.
\begin{figure}[t]
\begin{center}
\scalebox{1.1}{\includegraphics{xTOT.eps}}
\hspace{1.75cm}
\vspace{1.0cm}
\scalebox{1.1}{\includegraphics{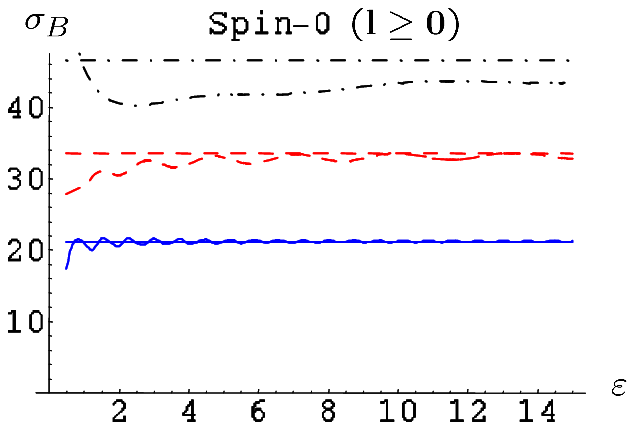}} 
\scalebox{1.1}{\includegraphics{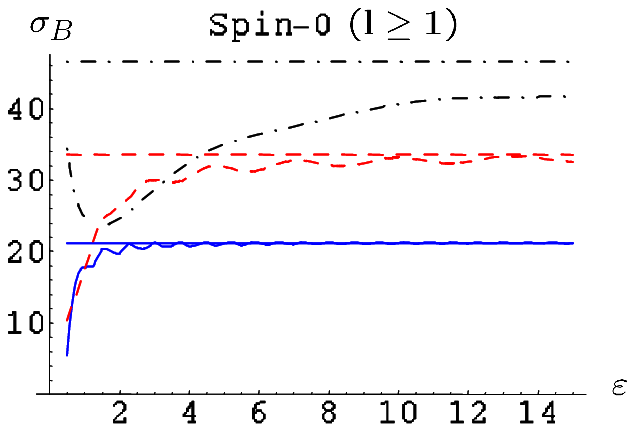}}
\hspace{1.75cm}
\scalebox{1.1}{\includegraphics{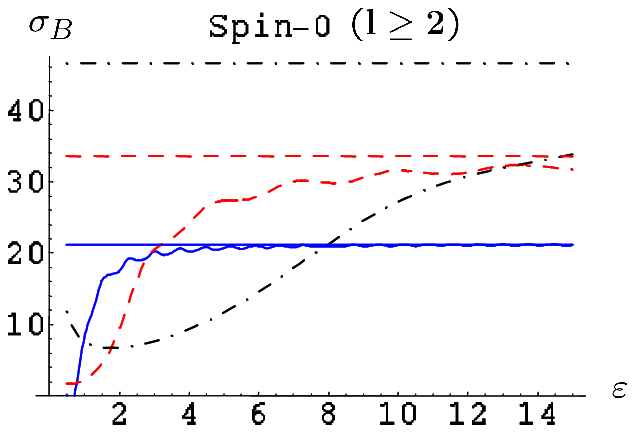}}
\vspace{0.25cm}
\end{center} 
\caption{Black hole cross-section for the graviton compared to a spin-zero field  as a function of $\varepsilon$, in units
of $(2M)^{n\over n-1}$. The horizontal lines correspond to the classical
cross-section. Blue (solid), red (dashed) and black (dot-dashed) curves correspond to $n=2,3$ and $4$ respectively. 
\label{bosex}}
\end{figure}

In Fig.~\ref{bosex} we compare the spin-zero cross-section, $\sigma_B$, with the total graviton cross-section. Although the classical cross-sections agree in the high-energy limit (asymptotically), at intermediate energies they do not agree due to differences in the partial wave sums. However, at lower energies we see that by taking the spin-0 cross-section from $l=1$ we obtain the best agreement with that for the graviton, see Fig.~\ref{bosex}. Note that for further comparison we also plot the spin-0 cross-section with the partial sum starting from $l=2$.

\newpage


\begin{thebibliography}{99}


\bibitem{GT}
S.~B.~Giddings and S.~Thomas,
Phys.\ Rev.\ D {\bf 65} (2002) 056010
[arXiv:hep-ph/0106219].

\bibitem{DL}
S.~Dimopoulos and G.~Landsberg,
Phys.\ Rev.\ Lett.\  {\bf 87} (2001) 161602
[arXiv:hep-ph/0106295].


\bibitem{HHBS}
S.~Hossenfelder, S.~Hofmann, M.~Bleicher and H.~Stoecker,
Phys.\ Rev.\ D {\bf 66} (2002) 101502
[arXiv:hep-ph/0109085].



\bibitem{Krev}
P.~Kanti,
Int.\ J.\ Mod.\ Phys.\ A {\bf 19} (2004) 4899
[arXiv:hep-ph/0402168].



\bibitem{FS1}
V.~P.~Frolov and D.~Stojkovic,
Phys.\ Rev.\ D {\bf 66} (2002) 084002
[arXiv:hep-th/0206046].

\bibitem{FS2}
V.~P.~Frolov and D.~Stojkovic,
Phys.\ Rev.\ Lett.\  {\bf 89} (2002) 151302
[arXiv:hep-th/0208102].

\bibitem{Stoj}
D.~Stojkovic,
Phys.\ Rev.\ Lett.\  {\bf 94} (2005) 011603
[arXiv:hep-ph/0409124].

\bibitem{FT}
A.~Flachi and T.~Tanaka,
  Phys.\ Rev.\ Lett.\  {\bf 95} (2005) 161302
  [arXiv:hep-th/0506145].



\bibitem{Card1}
V.~Cardoso, O.~J.~C.~Dias and J.~P.~S.~Lemos,
Phys.\ Rev.\ D {\bf 67} (2003) 064026
[arXiv:hep-th/0212168].

\bibitem{Konop}
R.~A.~Konoplya,
Phys.\ Rev.\ D {\bf 68} (2003) 024018
[arXiv:gr-qc/0303052].

\bibitem{Konop2}
R.~A.~Konoplya,
Phys.\ Rev.\ D {\bf 68} (2003) 124017
[arXiv:hep-th/0309030].
 

\bibitem{Card2}
V.~Cardoso, J.~P.~S.~Lemos and S.~Yoshida,
Phys.\ Rev.\ D {\bf 69} (2004) 044004
[arXiv:gr-qc/0309112].


\bibitem{Berti}
E.~Berti, M.~Cavaglia and L.~Gualtieri,
Phys.\ Rev.\ D {\bf 69} (2004) 124011
[arXiv:hep-th/0309203].


\bibitem{Chambers}
C.~M.~Chambers, W.~A.~Hiscock and B.~Taylor,
Phys.\ Rev.\ Lett.\  {\bf 78} (1997) 3249
[arXiv:gr-qc/9703018].


\bibitem{NYTM}
H.~Nomura, S.~Yoshida, M.~Tanabe and K.~i.~Maeda,
  Prog.\ Theor.\ Phys.\  {\bf 114} (2005) 707
  [arXiv:hep-th/0502179].


\bibitem{Harris:2005jx}
C.~M.~Harris and P.~Kanti,
arXiv:hep-th/0503010.

\bibitem{Duffy:2005ns}
G.~Duffy, C.~Harris, P.~Kanti and E.~Winstanley,
  JHEP {\bf 0509} (2005) 049
  [arXiv:hep-th/0507274].



\bibitem{Frolov}
V.~P.~Frolov and D.~Stojkovic,
Phys.\ Rev.\ D {\bf 67} (2003) 084004
[arXiv:gr-qc/0211055].


\bibitem{IOP2}
D.~Ida, K.~y.~Oda and S.~C.~Park,
Phys.\ Rev.\ D {\bf 67} (2003) 064025 [Erratum-ibid.\ D {\bf 69} (2004) 049901] [arXiv:hep-th/0212108];
ibid.\ D {\bf 71} (2005) 124039
[arXiv:hep-th/0503052].


\bibitem{KI}
H.~Kodama and A.~Ishibashi,
Prog.\ Theor.\ Phys.\  {\bf 110} (2003) 701
[arXiv:hep-th/0305147].

\bibitem{Zer}
F.~Zerilli, Phys. Rev. Lett. {\bf 24} (1970) 737; 
Phys. Rev. D {\bf 2} (1970) 2141.

\bibitem{RW} 
T.~Regge, and J.~Wheeler, 
Phys. Rev. {\bf 108} (1957) 1063.

\bibitem{Tang} 
F.~R.~Tangherlini, 
Nuovo Cimento {\bf 27}, 636 (1963). 

\bibitem{MP}
R.~C.~Myers and M.~J.~Perry,
Annals Phys.\  {\bf 172} (1986) 304.

\bibitem{CCG}
  V.~Cardoso, M.~Cavaglia and L.~Gualtieri,
  arXiv:hep-th/0512002; ibid. arXiv:hep-th/0512116.


\bibitem{Rubin}
  A.~Chodos and E.~Myers,
  Annals Phys.\  {\bf 156} (1984) 412;


  M.~A.~Rubin and C.~R.~Ordonez, 
  J.\ Math.\ Phys.\  {\bf 25} (1984) 2888; ibid.
  J.\ Math.\ Phys.\  {\bf 26} (1985) 65;
  
  I.~G.~Moss, ``Quantum theory, black holes and inflation,'' Wiley Press.

\bibitem{IW}
S.~Iyer and C.~M.~Will,
Phys.\ Rev.\ D {\bf 35} (1987) 3621.

\bibitem{Chandra}
S.~Chandrasekhar, ``The Mathematical Theory Of Black Holes,''
Clarendon Press, Oxford (1985).

\bibitem{PS}
E.~Poisson and M.~Sasaki,
Phys.\ Rev.\ D {\bf 51} (1995) 5753
[arXiv:gr-qc/9412027].

\bibitem{DeWitt}
B.~S.~DeWitt,
Phys.\ Rept.\  {\bf 19} (1975) 295.

\bibitem{EHM}
R.~Emparan, G.~T.~Horowitz and R.~C.~Myers,
Phys.\ Rev.\ Lett.\  {\bf 85} (2000) 499
[arXiv:hep-th/0003118].



\bibitem{KR}
P.~Kanti and J.~March-Russell,
Phys.\ Rev.\ D {\bf 66} (2002) 024023
[arXiv:hep-ph/0203223].

\bibitem{Unruh} 
W.~G.~Unruh,
Phys.\ Rev.\ D {\bf 14} (1976) 3251.

\bibitem{alstar}
A.~A.~Starobinsky, JETP 37, 28 (1973);
A.~A.~Starobinsky and S.M. Churilov, JETP 38, 1 (1973).


\bibitem{CJNS2}
A.~S.~Cornell, J.~Doukas, W.~Naylor and M.~Sasaki, in progress.

\bibitem{CSY}
V.~Cardoso, G.~Siopsis and S.~Yoshida,
Phys.\ Rev.\ D {\bf 71} (2005) 024019
[arXiv:hep-th/0412138].


\bibitem{KI2}
H.~Kodama and A.~Ishibashi,
Prog.\ Theor.\ Phys.\  {\bf 111} (2004) 29
[arXiv:hep-th/0308128].



\end{thebibliography}
\end{document}